%% file: main.tex
\documentclass[conference,10pt,bottom=0.95in]{IEEEtran}
\IEEEoverridecommandlockouts
\usepackage{graphicx,color}
\usepackage{algpseudocode}%

\usepackage{amsmath, amsthm, amsfonts, amssymb, amsbsy,nccmath}
\usepackage{algorithm}
\usepackage{enumerate}
\usepackage{lipsum}
\newtheorem{lemma}{Lemma}
\usepackage{textgreek}
\usepackage{comment}
\usepackage{textcomp}
\usepackage{amsfonts}

\usepackage[sort,compress]{cite}
\usepackage{epsfig}
\usepackage{epstopdf}
\usepackage{mathtools}
\usepackage{dsfont}
\usepackage{cleveref}
\usepackage{subfig}
\usepackage{stfloats}

\usepackage{epstopdf}

\theoremstyle{definition}

\usepackage[inline]{enumitem}   
\makeatletter
\newcommand{\inlineitem}[1][]{%
\ifnum\enit@type=\tw@
    {\descriptionlabel{#1}}
  \hspace{\labelsep}%
\else
  \ifnum\enit@type=\z@
       \refstepcounter{\@listctr}\fi
    \quad\@itemlabel\hspace{\labelsep}%
\fi}
\makeatother

\input{defines}

\newcommand{\bit}{\begin{itemize}}
\newcommand{\eit}{\end{itemize}}

\newcommand{\bms}{{\boldsymbol s}}

\DeclarePairedDelimiter\abs{\lvert}{\rvert}%

\usepackage{lipsum}

\usepackage{float}
\floatstyle{plaintop}
\restylefloat{table}

\makeatletter
\newcommand\fs@spaceruled{\def\@fs@cfont{\bfseries}\let\@fs@capt\floatc@ruled
  \def\@fs@pre{\vspace{0.5\baselineskip}\hrule height.8pt depth0pt \kern2pt}%
  \def\@fs@post{\kern1pt\hrule\relax}%
  \def\@fs@mid{\kern2pt\hrule\kern2pt}%
  \let\@fs@iftopcapt\iftrue}
  \newcommand\fs@betterruled{%
  \def\@fs@cfont{\bfseries}\let\@fs@capt\floatc@ruled
  \def\@fs@pre{\vspace*{5pt}\hrule height.8pt depth0pt \kern2pt}%
  \def\@fs@post{\kern2pt\hrule\relax}%
  \def\@fs@mid{\kern2pt\hrule\kern2pt}%
  \let\@fs@iftopcapt\iftrue}
\floatstyle{betterruled}
\restylefloat{algorithm}
\makeatother

\usepackage{footnote}

\begin{document}
\title{Next-Generation Sustainable Wireless Systems: Energy Efficiency Meets Environmental Impact\vspace{-2mm}}
\vspace{-4mm}
\author{ Christo Kurisummoottil Thomas\IEEEauthorrefmark{1}, Omar Hashash\IEEEauthorrefmark{2}, Kimia Ehsani\IEEEauthorrefmark{2}, and Walid Saad\IEEEauthorrefmark{2}\thanks{The research of Walid Saad was supported by the US National Science Foundation under Grant NSF CNS-2210254.} \\  \IEEEauthorrefmark{1} Department of Electrical and Computer Engineering, Worcester Polytechnic Institute, Worcester, MA, USA.\\  \IEEEauthorrefmark{2}Bradley Department of Electrical and Computer Engineering, Virginia Tech, Arlington, VA, USA.\\
Emails:cthomas2@wpi.edu,\{omarnh, kimiaehsani, walids\}@vt.edu \vspace{-0.4cm}}
\maketitle
\vspace{-4mm}
\begin{abstract}
Aligning with the global mandates pushing towards advanced technologies with reduced resource consumption and environmental impacts, the \emph{sustainability} of wireless networks becomes a significant concern in 6G systems. To address this concern, a native integration of sustainability into the operations of next-generation networks through novel designs and metrics is necessary. Nevertheless, existing wireless sustainability efforts remain limited to energy-efficient network designs which fail to capture the environmental impact of such systems.  
In this paper, a novel \emph{sustainability metric} is proposed that captures \emph{emissions per bit}, providing a rigorous measure of the environmental footprint associated with energy consumption in 6G networks. This metric also captures how energy, computing, and communication resource parameters influence the reduction of emissions per bit. Then, the problem of allocating the energy, computing and communication resources is posed as a multi-objective (MO) optimization problem. To solve the resulting non-convex problem, our framework leverages MO reinforcement learning (MORL) to maximize the novel sustainability metric alongside minimizing energy consumption and average delays in successfully delivering the data, all while adhering to constraints on energy resource capacity. 
 The proposed MORL methodology computes a global policy that achieves a Pareto-optimal tradeoff among multiple objectives, thereby balancing environmental sustainability with network performance. Simulation results show that the proposed approach reduces the average emissions per bit by around $26\%$ compared to state-of-the-art methods that do not explicitly integrate carbon emissions into their control objectives.
\end{abstract}

\vspace{-2mm}\section{Introduction}
\vspace{-1mm}


Ensuring access to sustainable energy resources is shortlisted as a top priority among the United Nation's sustainable development goals for $2030$~\cite{elavarasan2021envisioning}. Coinciding with the anticipated transition to 6G systems, \emph{sustainability} then becomes a crucial mandate that must be incorporated into the system design of next-generation wireless networks. Notably, the optimal design of 6G networks must be maximizing wireless quality of service (QoS) while reducing the energy consumption and its corresponding environmental footprint.  
Despite its anticipated role in the design of wireless networks, sustainability remains an abstract metric, that is challenging to quantify, and not well-defined in the prevalent literature ~\cite{PerinTNSM2021,LiuTWC2022,BinucciTGCN2023,QiIoTJ2024}. Although sustainability is directly linked to energy efficiency, carbon neutrality, and robust off-grid operations, \emph{capturing its precise definition remains necessary} to effectively analyze its impact in the realm of wireless networks.

Several prior works like \cite{PerinTNSM2021,QiIoTJ2024, LiuTWC2022,BinucciTGCN2023} attempted to capture sustainability in networks through energy efficient designs.
For instance, the works in \cite{PerinTNSM2021} 
 and \cite{LiuTWC2022} focus on the minimization of the energy cost in networks empowered with both renewable (e.g., solar) and non-renewable (e.g., grid) sources of energy while reducing the energy storage loss. Moreover,~\cite{BinucciTGCN2023} and \cite{QiIoTJ2024} minimize the energy consumption while considering QoS constraints in terms of latency and reliability. Although the works in \cite{LiuTWC2022, BinucciTGCN2023, PerinTNSM2021,QiIoTJ2024} may achieve energy efficiency, the proposed solutions do not inherently guarantee a low carbon emission. This is because, depending on the specific applications, achieving the desired QoS may require increasing computational resources. This, in turn, can lead to higher energy consumption and potentially higher carbon emissions. Indeed, it is possible to minimize the energy consumption while still contributing to a substantial carbon footprint which, in turn, defies carbon neutrality. Henceforth, it is essential to design a sustainability metric that can guarantee both energy efficiency and carbon neutrality by optimizing the allocation of energy resources between renewable and non-renewable sources.
 
\subsection{Related works}
Recently, a handful of works (e.g., \cite{RappaportArxiv2024,NGA2024,ITU2022} introduced new metrics to capture sustainability. For instance, the ``waste factor" is proposed in~\cite{RappaportArxiv2024} after a comprehensive analysis of the power utilization and energy waste at both the component and system levels in a source-to-sink communications system. However, despite the importance of this analysis for energy efficiency, this metric still fails to consider carbon emissions in its design. Furthermore, the work in \cite{NGA2024} proposes several metrics such as the ``carbon intensity of energy" that measures the carbon emissions associated with the energy consumption of a network element, in an attempt to reflect the environmental impact of its energy consumption. Nevertheless, this work falls short in providing a numerical formulation of such metric as a function of system parameters. \cite{ITU2022} introduces a metric called ``network carbon intensity (NCI)", defined as the ratio of total carbon emissions to total data traffic across the network. While NCI accounts for carbon emissions, it does not reflect the potential benefits of jointly optimizing computing, communication, and energy resources to reduce emissions; a limitation that this work seeks to overcome. A related work is \cite{SharafiRE2015} that proposes a ``renewable energy ratio (RER)" that captures the fraction of total energy load met by renewable sources; however, it focuses on energy management for buildings and lies outside the domain of wireless networks. Furthermore, the metrics introduced in~\cite{RappaportArxiv2024} and \cite{NGA2024} assess sustainability as an instantaneous measure rather than a long-term goal. 
Hence, these metrics neglect the evolution of energy resources over time and its impact on the overall sustainability of the network.
Consequently, these metrics cannot guarantee the long-term sustainability of wireless networks. 
The main contribution of this paper is a novel framework that guarantees long-term sustainability, energy efficiency, and QoS in a wireless system. In particular, we consider a multi-user downlink communication system with heterogeneous computing servers (CS) equipped with both CPU and GPU resources, and diverse energy sources.
To capture the sustainability of this system, we propose a \emph{novel sustainability metric} that considers: 1) efficient distribution of computing requirements for downlink signal processing operations among heterogeneous CS resources, 2) the optimal allocation of energy resources between renewable (i.e. limited in capacity) and non-renewable sources for required computational load, and 3) the optimization of the energy allocation in the wireless transmission to the users. Accordingly, we formulate a multi-objective (MO) optimization problem to maximize the sustainability and energy efficiency of the network, while minimizing the average delay under the capacity constraints on renewable energy sources. To solve this problem, we propose a MO reinforcement learning (MORL) framework. This MORL solution optimizes the energy efficiency, QoS, and long-term sustainability by computing parallel policies and fitting them into a single parametric distribution that  aligns with the local minima of the respective objective functions. Our simulation results demonstrate around $25-30\%$ reduction in carbon emissions while maintaining the QoS requirements compared to state-of-the-art systems that do not explicitly incorporate emissions per bit into wireless resource allocation.

\section{System Model and Problem Formulation}

\begin{figure}
    \centering
    \includegraphics[width=0.43\textwidth]{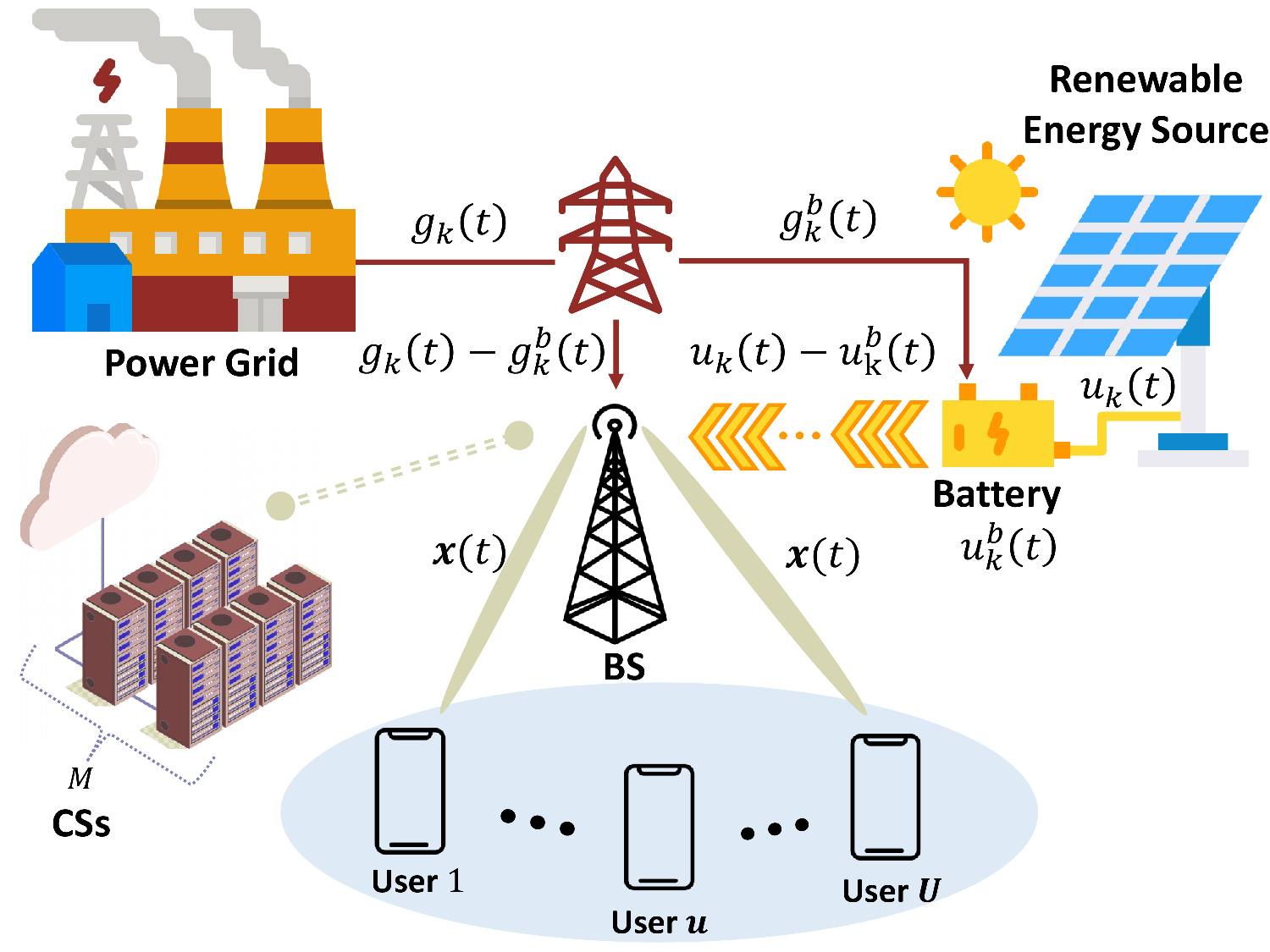}
    \vspace{-2mm}\caption{\small System model of the proposed sustainable wireless network.}
    \label{fig:System_Model}
    \vspace{-6mm}
\end{figure}

Consider a wireless network with a base station (BS) serving $U$ users in the downlink, as shown in Fig. \ref{fig:System_Model}. The system  timeline can be divided into timeslots of duration $T$ seconds (sec).  
At the beginning of each timeslot $t$, the BS transmits the data $x_{u}(t)$ to user $u$. Moreover, we assume the number of packets $\Lambda_u(t)$ arriving for transmission to user $u$ follows a Poisson process with a mean arrival rate of $\lambda_u$ packets/sec, where each packet has a fixed size of $L_u(t)$ bits. 
In the following, we quantify the BS computations required for transmitting $x_u(t)$ and the corresponding energy consumption.

To transmit the data $\bmx(t)=\left[x_1(t),\cdots,x_U(t)\right]$ of the $U$ users in timeslot $t$, signal processing operations at the BS can be concisely represented as: 
\vspace{-2mm}\beq
\phi(x) = \psi\left( \sum_{u=1}^{U} \nu_u(x_u) \right), \label{eq_hg_k}
\vspace{-2mm}\eeq
where $\nu_u$ encompasses user-specific operations such as source/channel coding, modulation, and precoding. Also, the function $\psi$ comprises common signal processing functions (associated  with 
 multiuser processing) in the 
radio-frequency chain. The computations involved in \eqref{eq_hg_k}, which may span multiple communication layers, vary in complexity, with some functions potentially requiring artificial intelligence. This variation in computational complexity motivates distributing the tasks across different servers or computing clusters to ensure efficient utilization of computing resources and energy \cite{PatelISCA2024}. As such, the computation and transmission loads as described by the operations in \eqref{eq_hg_k} are split among $M$ CSs. Subsequently, the processed packets are then transmitted to the $U$ users. We assume that \( M \leq U \).
Accordingly, the energy consumption model of this system comprises of the computing and communication costs during these operations. 

In essence, the energy supply for each CS includes non-renewable (i.e., power grid) and
renewable energy sources (e.g., solar power). Moreover, each CS $k$ is equipped with a battery than can operate either as an energy source or sink. As such, we denote $d_k(t)$ as the energy consumed by CS $k$ from its battery at time slot $t$. In addition, $u_k(t)$ denotes the energy extracted from a shared renewable source (across all CSs) at CS $k$ during time slot $t$. Here, $u_k^b(t)$ is the renewable energy utilized for charging the battery and the remaining $u_k(t)-u_k^b(t)$ is the renewable energy immediately utilized by the CS for its operations. 
Similarly, we define $g_k(t)$ as the energy drawn from the power grid by CS $k$ during time slot $t$, $g_k^b(t)$ as the portion of the energy allocated to charging the batteries, and the remaining $g_k(t) - g_k^b(t)$ as the energy used immediately.
The battery level $b_k(t)$ evolves according to a standard model~\cite{CecchinatoICASSP2020} and can be described as: \vspace{-0.2cm}
\beq
b_k(t) = \eta_k\left(b_k(t-1) - d_k(t)\right) + \mu_k\left(u_k^b(t) + g_k^b(t)\right), \vspace{-0.1cm}
\eeq
where the parameter $\eta_k \in (0, 1]$ represents the battery's self-discharging behavior and $\mu_k \in (0, 1]$ accounts for losses incurred during the charging process. Notably, the coefficients $\eta_k$ and $\mu_k$ can be either calculated using the storage loss models which capture the capacitor losses in the battery~\cite{BiasonICNC2016} or can be learned using neural network models.


\subsection{Computing model}
According to standard models~\cite{ZengTWC2021}, the total workload $W_k(t)$ for the CS $k$ is
a fraction of the total load $N_{f}\sum\limits_u L_u(t)$,
where $N_{f}$ denotes the number of floating point operations
(FLOPs) needed for processing each bit. We assume that each CS contains distinct computing components of varying processing capabilities, such as GPUs and CPUs. Furthermore, the CPU computing speed can be defined as $f_k = n_k \times f_k^C$, where $n_k$ is the number of CPU FLOPs/cycle. Similarly, the GPU computing speed is $f_k^{\prime} = n_k^{\prime} f_k^G$, where $n_k^{\prime}$ is the number of GPU FLOPs/cycle. Moreover, $f_k^C$ and $f_k^G$ (cycles/sec) are the clock frequency
of the CPU and GPU at CS $k$, respectively. To facilitate workload partitioning at CS $k$, the computations between the CPU and GPU are split according to a partitioning factor $\kappa_k$.
Thus, the computation delay at each CS becomes: \vspace{-0.2cm}
\beq 
\begin{aligned}
\tau_{o,k}(t) = \max\left\{\underbrace{\kappa_k\frac{W_k(t)}{f_k(t)}}_{\tau_{o,k}^C(t)},\underbrace{(1-\kappa_k)\frac{W_k(t)}{f_k^{\prime}(t)}}_{\tau_{o,k}^G(t)}\right\}.
\end{aligned}
\vspace{-0.1cm}
\eeq 

To quantify the energy consumed during signal processing operations of \eqref{eq_hg_k}, 
the energy consumed by GPU/CPU computations at CS $k$ can be written as follows~\cite{TsuzukuICT2015_1}:\vspace{-0.10cm}
\beq\begin{aligned}
E_{s,k}(t) = \left(\alpha_{k} A_s f_{k}^s(t)^3 + \beta_{k} B_s + C_s f_{k}^s(t)\right)\tau_{o,k}^s(t), 
\label{eq_E_GPU}
\end{aligned}
\vspace{-0.05cm}
\eeq
where $s\in \{G,C\}$, $\alpha_k \!\in\! [0,1]$ is the parameter (representing compute intensity) and $\beta_k \in [0,1]$ is a data-dependent application parameter. In addition, $A_s$ and $B_s$ are GPU/CPU architecture dependent parameters and $C_s$ corresponds to power losses due to device impairments.
Notably, the
first term in~\eqref{eq_E_GPU} is the power consumed by the GPU/CPU
cores for computation, the second term is
the power consumed by the GPU/CPU memory, and the third term corresponds to the static power consumption. Here, we assume that all the computations by the GPU/CPU must be completed within a slot duration $T$ to ensure an uninterrupted user experience. 

\vspace{-0.2cm}
\subsection{Communication model}
Considering a dynamic wireless channel and a capacity-achieving transmission system over a bandwidth $B$, the downlink signal received at user $u$ can be formulated as:
\vspace{-1mm}\beq
y_u(t) = h_{u}(t)x_u(t) + \sum\limits_{i\neq u}h_{u}(t) x_i(t) + n_u(t),
\vspace{-1mm}\eeq
where $h_{u}(t)$ is the effective downlink channel coefficient, $x_u(t) \sim \mN(0,p_u(t))$ is the transmit signal with the transmit power $p_u(t)$ for user $u$,  $n_u(t) \sim \mN(0,N_0B)$ is assumed identical and independent across the time slots, and $N_0$ is the noise power
spectral density at the receiver side. Accordingly, the downlink rate for user $u$ follows the standard expression:
$
R_u(t) \!=\! B\log_2 \bigl(1+ \frac{p_u(t)\left(\abs{h_{u}(t)}^2\right)}{(\abs{h_{u}(t)}^2\sum\limits_{i\neq k}p_i(t)+N_0B)}\bigr).$
By inverting $R_u(t)$, we capture the transmission energy for $x_u(t)$ as:
\beq\begin{aligned}
E_{\textrm{tr},u}(t) = \frac{ (\abs{h_{u}(t)}^2\sum\limits_{i\neq u}p_i(t)+N_0B)T}{\abs{h_{u}(t)}^2 }\left(e^{\frac{R_u(t)\ln 2}{B}}-1\right).
\label{eq_E_Comm}
\end{aligned}
\eeq
Therefore, by combining \eqref{eq_E_GPU}, and \eqref{eq_E_Comm}, the total energy consumed at CS $k$ is derived as follows:
\vspace{-1mm}\beq
E_k^{\textrm{tot}}(t) = \underbrace{\bigl[\sum\limits_{u\in \mM_k}E_{\textrm{tr},u}(t)\bigr]}_{\textrm{Communication}} + \underbrace{E_{G,k}(t)+E_{C,k}(t)}_{\textrm{Computing}}, 
\label{eq_Ek_tot}
\vspace{1mm}\eeq
where $\mM_k$ is the set of user computations performed at CS $k$ and is assumed to be fixed. The total energy consumed is extracted from both renewable and non-renewable sources, leading to the  constraint:
\vspace{-2mm}\beq
E_k^{\textrm{tot}}(t) = u_k(t) - u_k^b(t) + g_k(t) - g_k^b(t) + d_k(t).
\label{eq_Ek}
\vspace{-1mm}\eeq
To capture the  transmission delay (due to $M$ transmission queues), we derive the number of packets that can be transmitted during the $t$-th time-slot as:
$N_k(t) = \lfloor \frac{T \sum\limits_{u\in \mM_k}R_u(t)}{\abs{M_k}L_k(t)}  \rfloor.$
Moreover, the evolution of the queue size $\Omega_k
(t)$ corresponding to arrival of packets for $x_u(t), \forall u\in \mM_k$ is formulated as:
\beq
\Omega_k(t+1) = \max(0, \Omega_k(t) - N_k(t)) + \Lambda_k(t).
\eeq
Using Little's law \cite{LittleOR1961}, the average long-term queuing delay, gets written as:
$
{\tau}_{q,k} = \lim_{T\rightarrow \infty} \frac{1}{T}\sum\limits_{t=1}^T\mbE\{\frac{\Omega_k(t)}{{\lambda}_k}\}.$
Further, the average total delay encountered by any packet $x_k(t)$ becomes:
\beq
{\tau}_k(t) = \underbrace{{\tau}_{q,k}(t)}_{\textrm{Queuing delay}} + \underbrace{\sum\limits_{u\in \mM_k}\frac{L_u(t)}{\bar{R}_u(t)}}_{\textrm{Communication delay}} + \underbrace{{\tau}_{o,k}(t)}_{\textrm{Computing delay}}.
\eeq

 \subsection{Sustainability metric}
Since energy efficiency does not necessarily imply sustainability, our primary goal is then to achieve both carbon emission reduction and energy efficiency.
To this end, it is necessary to optimize the energy allocation between renewable and non-renewable resources in the network.
Thus, the carbon emissions due to the consumed energy can be written as:
\beq
c_k(t) = w_{u}(u_k(t) - u_k^{b}(t)) + w_{g}(g_k(t)-g_k^{b}(t)) + w_{d}d_k(t),
\label{eq_emissions}
\eeq
where the weights $w_u,w_g,$ and $w_d$ are the emission factors \cite{MarrassoECM2019} for converting the energy consumed to the amount of $\mathrm{CO_{2}}$ emitted (amount of carbon emissions per Joule of energy) \cite{VoumikEnergies2023}. The emission factors are assumed to be unknown, since they are not constant and can vary due to the type of fuel mixed to generate electricity, time of the day, location and the technology used \cite{MarrassoECM2019}. 
Further, we introduce the \emph{carbon emissions per bit measure}, which is used to measure the sustainability of a wireless system that uses a combination of computational resources optimization, renewable and non-renewable sources to reduce the carbon emissions compared to conventional networks that uses non-renewable sources at the CSs. The carbon emissions per bit can be written as:
$
C^b_{k}(t) = \frac{c_k(t)}{T \sum\limits_{u\in\mM_k}R_u(t)}.$
Using the aforementioned metrics, we define the sustainability measure as follows:
\vspace{-2mm}\beq
S(t) = \left(1- \frac{ \sum\limits_{k=1}^{M} C^b_{k}(t)}{C_\psi^b(t)}\right)^2,
\vspace{-1mm}\eeq
where $C_\psi^b(t) =  w_g \frac{\sum\limits_{k=1}^ME_k^{\textrm{tot}}(t)}{T\sum\limits_{u=1}^U R_u(t)}$ is carbon emissions per bit when the whole energy requirements are provided by a non-renewable source.
Next, we formulate an optimization problem that aims to optimize the network sustainability and energy efficiency while meeting the user QoS requirements.

\vspace{-1mm}\subsection{Problem formulation}
\vspace{-1mm}

Our goal is to \emph{optimize the long-term network sustainability, while simultaneously meeting QoS requirements (i.e., delay) and guaranteeing energy efficiency}\footnote{\vspace{-0.5mm}We emphasize that it is straightforward to extend this framework with additional QoS metrics such as reliability and throughput in future works.}. This problem can be formulated as a MO optimization problem, as follows:
\vspace{-1mm}\begin{subequations}
\vspace{-1mm}\begin{align}
\mP_1: \max\limits_{\{\bma_k(t)\}_{\forall k}} \,\,\, \,\,& \!\!\!\!\Bigl[\hspace{-0mm}\!\underbrace{\lim_{T\rightarrow \infty}\frac{1}{T}\sum_{r=1}^TS(t+r)}_{\mbox{\small long-term sustainability }}\!, \,\!\nonumber\\ \vspace{-1mm} & \,\,\,\,\,\,\,\,\,\,\,\,\,\,\,\,\,\,\,\,\,\,\,\,\,\,\,\,\,-\!\lim_{T\rightarrow \infty}\frac{1}{T}\sum\limits_{r=1}^T\frac{\sum\limits_{k}E_k^{\textrm{tot}}(t+r)}{T\sum\limits_{u} R_u(t+r)}, \nonumber\\ \vspace{-1mm} & \,\,\,\,\,\,\,\,\,\,\,\,\,\,\,\,\,\,\,\,\,\,\,\,\,\,\,\,\,-\lim_{T\rightarrow \infty}\sum\limits_k\frac{1}{T}\sum\limits_{r=1}^T \tau_k(t+r)\Bigr] \\ \vspace{-6mm}
\mbox{s.t.}\,\,\,\,\,\,\,\,\,\,\,\, 
&0 \leq \kappa_k(t) \leq 1, \forall k,t,\\ & d_k(t) \geq 0, u_k(t) \geq 0, u_k^b(t) \geq 0, \nonumber\\ &g_k(t) \geq 0,g_k^b(t) \geq 0 \,\,\,\forall k, \forall t, \label{nonneg}\\
& b_k(t) \leq b_k^{\textrm{max}}, \,\,\, d_k(t) \leq b_k(t-1), \forall k, \label{battery_level} \\ 
  \mD_1:\,\,\,\,\,\,\,\,\,\,\,\,  &\sum\limits_ku_k(t) \leq U^{\textrm{max}}, \label{eq_renew_capacity} \\
\mD_2: \,\,\,\,\,\,\,\,\,\,\,\, & \!\!\!\!\!\!\!\!\!\,\,\eqref{eq_Ek_tot}\,\, \mbox{and}\,\, \eqref{eq_Ek} , \forall k, t \label{eq_energy_split_user} \\
 \vspace{-2mm}
 \mD_3: \,\,\,\,\,\,\,\,\,\,\,\, &\sum\limits_kp_k(t) \leq P, \forall t,
\vspace{-7mm}
\end{align}
\label{eq_P1}
\end{subequations}
\noindent where $\bma_k(t) =\left[ \{ u_k^b(t), g_k^b(t),\{p_u(t)\}_{u\in \mM_k},f_k(t),\kappa_k(t)\}\right]$ captures the optimization variables. Here, \eqref{nonneg} represents the non-negativity constraints on all the energy variables and \eqref{battery_level} defines the constraints on battery capacity and amount of energy that can be drained from it. Moreover, \eqref{eq_renew_capacity} is the renewable energy capacity limit across all CSs. \eqref{eq_energy_split_user} is the constraint that the total consumed energy at each CS should be split between renewable and non-renewable sources. In addition, $P$ is the maximum transmit power across all users. 

The problem in $\mP_1$ is challenging due to the dependencies between the objectives. In particular, a tradeoff between sustainability and QoS of the users rises upon optimizing $f_{k}$ and $p_k$. Increasing $f_k$ minimizes latency, however, energy increases as per \eqref{eq_E_GPU}. While increasing $p_k$ can enhance the transmission rate, it incurs an additional energy consumption. Consequently, this degrades the overall sustainability of the network. Similarly, while allocating additional processing energy requirements to a renewable source (i.e, higher values of $u_k(t)-u_k^b(t), d_k(t)$) enhances sustainability, it increases the transmission latency due to the additional time needed to recharge the batteries. Thus, it is challenging to find a single global optimum for $\mP_1$. 
In this regard, we solve \eqref{eq_P1} using multi-objective reinforcement learning (MORL) to optimize resource allocation, while ensuring an optimal tradeoff among long-term sustainability, energy efficiency, and average delay.

\vspace{-1mm}\section{Proposed Solution: MORL for Pareto Optimal Sustainable Network Design}\vspace{-1mm}

\begin{figure}[t]
\vspace{-3mm}
\begin{algorithm}[H]
\caption{\small Proposed algorithm}
\label{algo:consecutive_best_response}
\begin{algorithmic}[1]\small
\vspace{-2mm}
\State \textbf{Given:} Initialize the policy and the approximate model $\widehat{P}$ \;
\For{each epoch}
     \State Collect data with $\pi(\bma_t\mid\bms_t,\mV)$ in real environment: $\mD\leftarrow \mD \cup \{\bms_{t+r},\bma_t,\bms_{t+r+1}\}_{\forall r \in [0,T]}$\;
   \State Optimize $\widehat{P}$ for multi-step prediction under the current policy $\pi:\widehat{P}\leftarrow \argmax\limits_{\widehat{P}} F $\;, with $\widehat{P}$ represented using an LSTM.
    \State Optimize $\pi(\bma_t\!\!\mid\!\!\bms_t,\mV)$ under MORL 
 as discussed in ~\ref{MORL}.\;
\EndFor
\end{algorithmic}
\label{Algo1}
\end{algorithm} 
\vspace{-10mm}
\end{figure}
To solve \eqref{eq_P1}, we consider a two-stage procedure as described in Algorithm~\ref{Algo1}. In the first stage, drawing inspiration from \cite{PerinTNSM2021}, we  track the evolution of carbon emissions (note that  \cite{PerinTNSM2021} focuses on tracking energy consumption instead) described in \eqref{eq_emissions} over a specified time horizon of $T$ slots. Such predictive networks enable more effective planning of energy, computing, and communication resources, as opposed to policies based solely on the state of each individual time slot. Using the predictive network, we evaluate the objective functions in \eqref{eq_P1}. In the second stage, a MORL solution is applied to compute $\bma_k(t), \forall k$) by optimizing \eqref{eq_P1}. 

\begin{figure*}\vspace{-0mm}\small\beq
\begin{aligned}
\small Q_1(\bms_t;\bma_t) &= \mbE_{\pi}\left[\frac{1}{n}\sum\limits_{r=0}^{n}S(t+r)\right]  - \lambda_1\left(\sum\limits_{t\in\mN}\sum\limits_k(u_k(t)) - U^{\textrm{max}}\right)  - \sum\limits_k\lambda_{2,k} \sum\limits_{t\in\mN}\left(u_k(t)-u_k^b(t) +\!g_k(t)-g_k^b(t)+ \!d_k(t) - \!E_k^{\textrm{tot}}(t)\right) ,\\ \vspace{-1mm} 
\small Q_2(\bms_t;\bma_t) &= --\lim_{T\rightarrow \infty}\frac{1}{T}\sum\limits_{t=1}^T\frac{\sum\limits_{k}E_k^{\textrm{tot}}(t)}{T\sum\limits_k R_k(t)} - \lambda_3\left(\sum\limits_{t\in\mN}\sum\limits_kp_k(t) - P\right) -\lambda_4 \left(\sum\limits_{t\in\mN} Tp_k(t) + E_{G,k}(t)+ E_{C,k}(t) - E_k^{\textrm{tot}}(t)\right),\\ \vspace{-1mm}
\small Q_3(\bms_t;\bma_t) &= -\mbE_{\pi}\left[\sum\limits_{r=0}^{n}\tau_k(t)\right] - \lambda_3\left(\sum\limits_{t\in\mN}\sum\limits_kp_k(t) - P\right) -\lambda_4 \left(\sum\limits_{t\in\mN} Tp_k(t) + E_{G,k}(t)+ E_{C,k}(t) - E_k^{\textrm{tot}}(t)\right).
\label{eq_Lagrangian}\end{aligned}
\eeq\vspace{-9mm}
\end{figure*}

\vspace{-2mm}\subsection{MORL framework}
\label{MORL}

We formulate the problem $\mP_1$ as a MORL, where the states are defined as  emissions per bit $\bms_t = \left(\bmC_1^b(t),\cdots,\bmC_{M}^b(t)\right)$. We assume that the channel $h_{k}$, and the bandwidth allocated to the users $B$ are known factors and are denoted as the set $\mV$. The vector of actions is defined as $\bma_t = \left(\bma_1(t),\cdots,\bma_K(t)\right)$. Here, the state transitions are deterministic and follow \eqref{eq_emissions}. Given the unknown emission factors, we propose to predict the future states across $T$ timeslots using long-short-term memory networks (LSTM) by optimizing the prediction error. We denote the predicted states as the environment model $\widehat{P}$.  The considered Markov decision process (MDP) here is an MO-MDP, which is an
MDP in which the reward function is composed of three components, as defined in \eqref{eq_P1}. Here, we define the policy using the probability distribution $\pi(\bma_t\!\mid\!\!\bms_t,\mV)$. The resolution to a MO-MDP involves a collection of policies referred to as a ``coverage set." 

\subsubsection{Policy improvement for multi-objective optimization}

The problem in \eqref{eq_P1} is a constrained MDP. Hence, we define the Q-functions for the three objectives using the augmented Lagrangian approach as \eqref{eq_Lagrangian}.
Here, we consider a set of binary random variables $Z_r$ that indicates a policy improvement event with respect to
objective $Q_r(\boldsymbol{s}_t;\boldsymbol{a}_t)$. $Z_{r,t} = 1$ indicates that the policy has improved for objective $r$ at time $t$, while $Z_{r,t} = 0$ indicates
that it has not. We seek policy $\pi(\bma_t\mid\bms_t,\mV)$ that maximize the following marginal likelihood, given preference tradeoff coefficients $\{\zeta_r\}:$
\begin{equation}
\max\limits_{\pi(\bma_t\mid\bms_t,\mV)} \mathbb{E}_{\mu} \log p \left(\{Z_{r,t}=1\}_{r=1}^{3}, \forall t\in \mN\mid \{\zeta_{r}\}_{r=1}^3,\bms_t\right),
\end{equation}
where $\mu = p(\bms_t\mid\bma_t)$.
Assuming that $Z_r$'s are all independent, we simplify the likelihood using the following general model \cite{AbdolmalekiArxiv2021}:
$
F = \mathbb{E}_{\mu} \sum\limits_{r=1}^{3} \log p \left(Z_{r,t},\forall t\in\mN\mid \zeta_r,\bms_t\right)^{\zeta_r}.$
Since this posterior is intractable, we introduce variational distributions $q_r(\boldsymbol{a}_t\mid\boldsymbol{s}_t)$. To compute these approximate distributions, we resort to variational Bayesian inference methods, which optimizes a lower bound of the likelihood function $F$ and the resulting alternating optimization can be written as \eqref{eq_P}. We have followed similar derivation as in 
\cite{AbdolmalekiArxiv2021} but extended to a wireless network model and hence, we skip the details here. We solve this MORL problem using an alternating optimization method. In the first step of this procedure, we compute $q_r$ as follows.
\begin{figure*}\beq
\left[\pi^{\ast}(\bma_t\mid\bms_t,\mV),\\ q_r^{\ast}(\bma_t\mid \bms_t)\right]=\max\limits_{\substack{\pi(\bma_t\mid\bms_t,\mV),\\ q_r(\bma_t\mid \bms_t)}}-\sum\limits_{r=1}^3\zeta_r\mathbb{E}_{\mu}\text{KL}(q_r(\boldsymbol{a}_t\mid\boldsymbol{s}_t)\mid \mid p(Z_{r,t}\mid\boldsymbol{s}_t,\boldsymbol{a}_t)\pi_i(\boldsymbol{a}_t\mid\boldsymbol{s}_t)).
\label{eq_P}\eeq\vspace{-6mm}
\end{figure*}

\vspace{-2mm}\begin{equation}
\vspace{-1mm}\begin{aligned}
&\max\limits_{q_r}\! -\!\sum\limits_{r=1}^3\zeta_r\mathbb{E}_{\mu}\text{KL}(q_r(\boldsymbol{a}_t\mid\boldsymbol{s}_t)\mid \mid p(Z_{r,t}\mid\boldsymbol{s}_t,\boldsymbol{a}_t)\pi_i(\boldsymbol{a}_t\mid\boldsymbol{s}_t)),
\label{eq_q_r_min}
\end{aligned}
\vspace{-0mm}\end{equation} 
where, $KL(q\mid\mid p)$ is the Kullback-Leibler divergence (KLD) measure between distributions $q$ and $p.$
In the second step, we maximize the lower bound $L$ with respect to $\pi(\bma_t\mid\bms_t,\mV)$, holding $q_r$ fixed to the solution of \eqref{eq_q_r_min}. This gets written as:
\vspace{-1mm}\begin{equation}
\begin{aligned}
\max\limits_{\pi(\bma_t\mid\bms_t,\mV)}  -\sum\limits_{r=1}^3\zeta_r\mathbb{E}_{\mu} \text{KL}(q_r(\boldsymbol{a}_t\mid\boldsymbol{s}_t)\mid\mid\pi(\boldsymbol{a}_t\mid\boldsymbol{s}_t,\mV)).
\label{eq_global_policy}
\end{aligned}
\end{equation}

We first notice that the objective \eqref{eq_q_r_min} is a weighted sum of three independent terms that can all be optimized separately. Hence, 
$\min\limits_{q_k} \mathbb{E}_{\mu}\text{KL}(q_r(\boldsymbol{a}_t\mid\boldsymbol{s}_t)\mid\mid p(Z_{r,t}\mid\boldsymbol{s}_t,\boldsymbol{a}_t)\pi(\boldsymbol{a}_t\mid\boldsymbol{s}_t,\mV))$
leads to the following solution:
\vspace{-2mm}\begin{equation}
\begin{aligned}
&q_r(\boldsymbol{a}_t\mid\boldsymbol{s}_t)\! \\ &= \!\frac{\sum\limits_{t=1}^np(Z_{r,t}\mid\boldsymbol{s}_t,\boldsymbol{a}_t)\pi(\boldsymbol{a}_t\mid\boldsymbol{s}_t,\mV)e^{-\sum\limits_{i}\lambda_i\mathbb{I}(\mD_i)}}{\int \sum\limits_{t=1}^np(Z_{r,t}\mid\boldsymbol{s}_t,\boldsymbol{a}_t)\pi(\boldsymbol{a}_t\mid\boldsymbol{s}_t,\mV)e^{-\sum\limits_{i}\lambda_i\mathbb{I}(\mD_i)}d\boldsymbol{a}} ,
\end{aligned}
\end{equation}
where $\mathbb{I}(\mD_i)$ is the indicator function which is true when the respective constraint in \eqref{eq_P1} is satisfied. The Lagrange multipliers $\lambda_i$ can be computed using bisection.
This nonparametric distribution adjusts the action weights according to the probability of improvement, $p(Z_{r,t}\mid\boldsymbol{s}_t,\boldsymbol{a}_t)$, which is modeled following the standard practice in RL literature:
\vspace{-0mm}\begin{equation}
p(Z_{r,t}\mid\boldsymbol{s}_t,\boldsymbol{a}_t) \propto \sum\limits_{t=1}^n\frac{Q_r(\boldsymbol{s}_t,\boldsymbol{a}_t)}{\rho_k}.
\vspace{-0mm}\end{equation}
Here, $\rho_r$ is a temperature parameter that regulates how greedy the solution $q_r(\bma_t\mid \bms_t)$ is with respect to its objective.

\vspace{-2mm}\begin{lemma}
    The global policy that minimizes \eqref{eq_global_policy} can be written as the weighted average distribution 
    \vspace{-2mm}\beq\pi(\bma_t\mid\bms_t,\mV) = \sum\limits_{r=1}^3\zeta_rq_r(\bma_t\mid\bms_t).
    \eeq \vspace{-2mm}
\end{lemma}
\vspace{-4mm}\begin{IEEEproof}
    The proof follows directly by expanding the KLD terms in \eqref{eq_global_policy}. We omit the details due to space constraints.
\end{IEEEproof}
Given KLD is a convex function, the resulting alternating optimization between the local policies $q_r$ and the global policy $\pi(\bma_t\mid\bms_t,\mV)$ leads to a local optimum solution.

\section{Simulation Results and Analysis}
\label{Simulation}

\begin{table}[t]
    \centering \small
    \vspace{1mm}\begin{tabular}{c|c}
    \hline
     Parameter & Value \\[0.5ex] 
      \hline\hline
        $\eta_i $ & $0.9999$ \\
           \hline
         $\mu_i$ &  $0.900$\\
            \hline
            $U^{\textrm{max}} $ & $1.00$kWh\\
            \hline
            $b_k^{\textrm{max}}$ & $2$kWh\\ \hline
            $P $ & $1$W \\
            \hline
            $U$ & $10$ \\
            \hline
            $M$ & $4$\\
            \hline
            $B$ & $5$MHz\\
            \hline
            $f_k^C$ & $[2,3]$ GHz \\ \hline
            $f_k^G$ & $[10,12]$ GHz \\ \hline
            $N_f$ & $[5,15]$ MFLOPS \\ \hline
    \end{tabular}
 \vspace{-2mm}   \caption{Simulation Parameters }
    \label{tab:my_label}
\vspace{-5mm}\end{table}

For our simulations, we consider $h_k(t)$ to be generated as a real Gaussian with mean $\mu = 0$ and variance $\sigma^2$ such $0.8 \leq \sigma^2 \leq 1.2$. The received SNR is defined as: $SNR_k = E{|h_k(t)|^2}/(N_0B_k)$ and is assumed to have a constant value of $10$ dB. The mean number of packets arriving at the start of any time slot is randomly sampled from a uniform distribution in the interval $[1,30]$.  The rest of the simulations parameters are described in Table~\ref{tab:my_label}. Our benchmarks for comparison include an: a) \emph{MORL with scalarization} scheme that converts the MO problem into a single-objective problem via a weighted combination of multiple objectives,  b) \emph{MORL with energy efficiency optimization} that does not incorporate sustainability into its energy minimization objective, and c) a state of the art  sustainability measure called \emph{RER} being used instead of the proposed metric \cite{SharafiRE2015}.

\begin{figure}[t]
    \centering
\includegraphics[width=0.46\textwidth,height=0.22\textwidth]{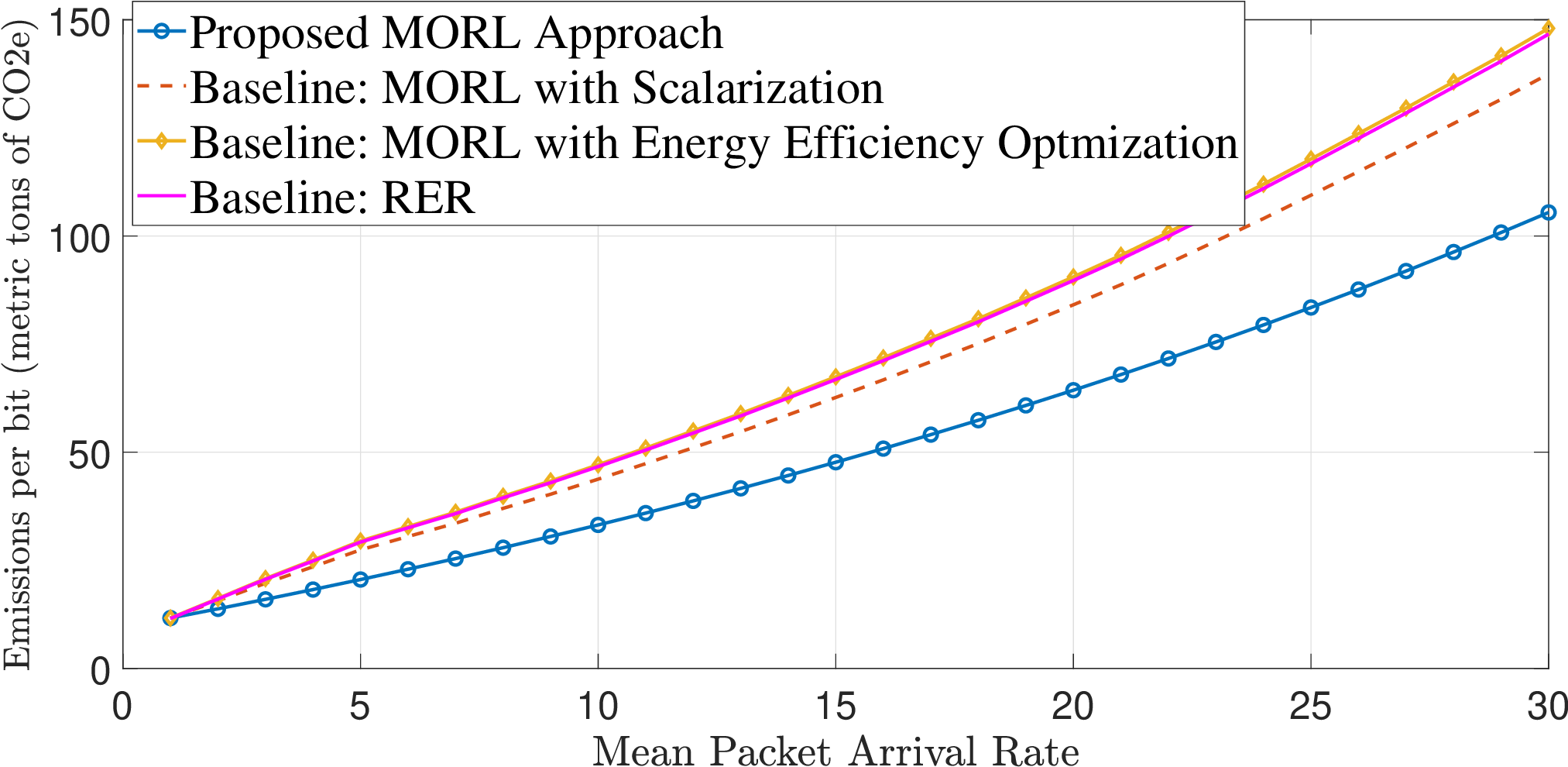}
\vspace{-2mm}\caption{\small Emissions per bit versus mean packet arrival rate}
    \label{fig:emissions_vs_packet} 
    \vspace{-4mm}
\end{figure}

Fig. \ref{fig:emissions_vs_packet} shows that our proposed  MORL approach can outperform the energy efficiency optimization benchmark as it factors in carbon emissions into the sustainability measure, as shown in \eqref{eq_emissions}.
In fact, our proposed solution can achieve a $26\%$ lower average carbon emissions for varying packet arrival rates in comparison to the baselines, while achieving the same energy efficiency and average delay values. 

\begin{figure}[t]
    \centering
\includegraphics[width=0.46\textwidth,height=0.22\textwidth]{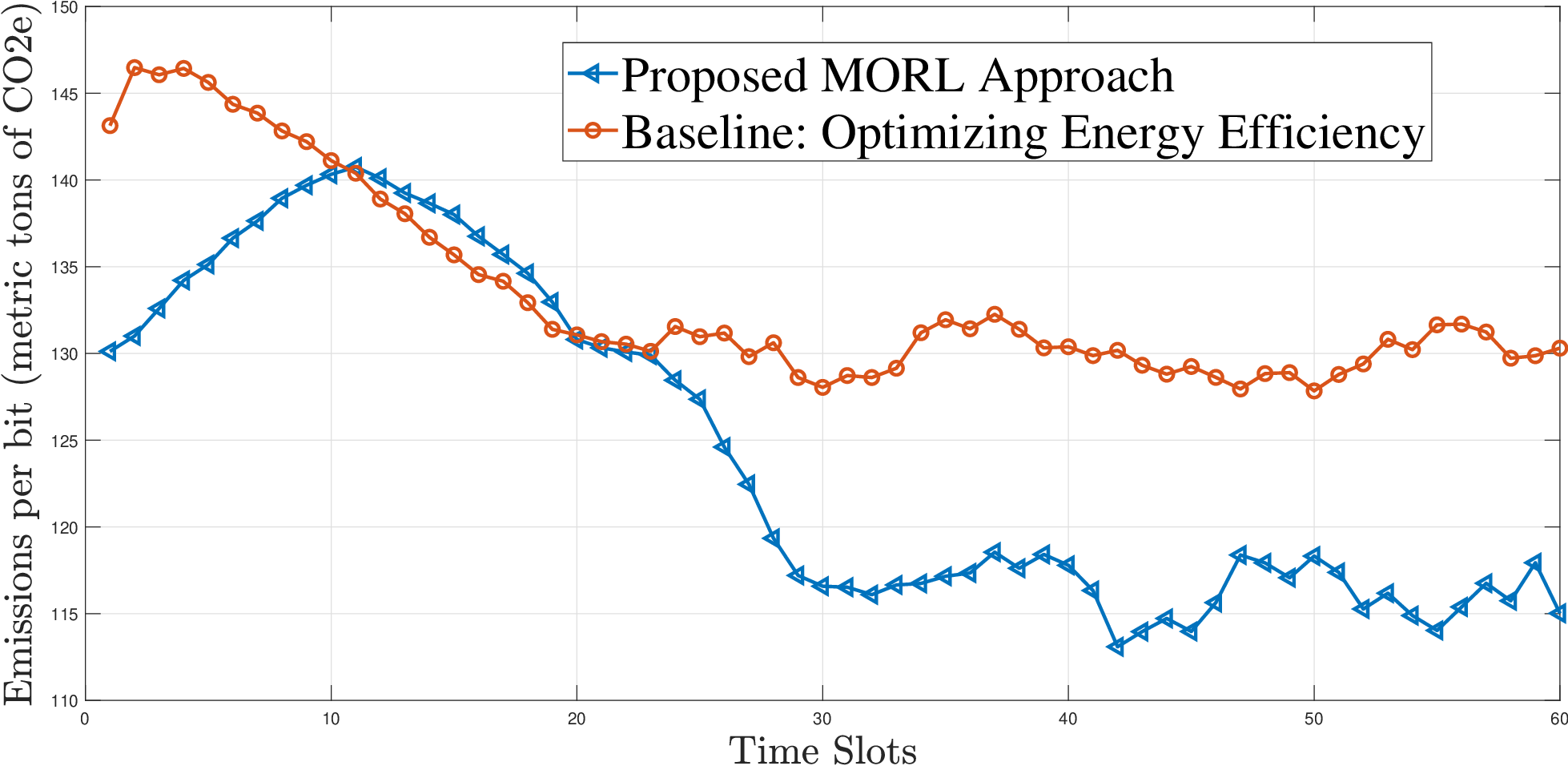}
\vspace{-2mm}\caption{\small Emissions per bit vs time for a fixed packet arrival rate.}
    \label{fig:emissions_vs_time} 
    \vspace{-7mm}
\end{figure}

Fig. \ref{fig:emissions_vs_time} shows that our MORL framework reduces the overall $\mathrm{CO_{2}}$ emissions per bit over multiple timeslots as compared to the energy efficiency benchmark. Therefore, our MORL framework can ensure long term sustainability as its predictive measures favor average reductions over time rather than individual time-slots. Proposed MORL approach shows almost $11\%$ reduction in emissions per bit at convergence compared to baseline at an average delay of $0.2$ ms.

\begin{figure}[t]
    \centering
\includegraphics[width=0.46\textwidth,height=0.22\textwidth]{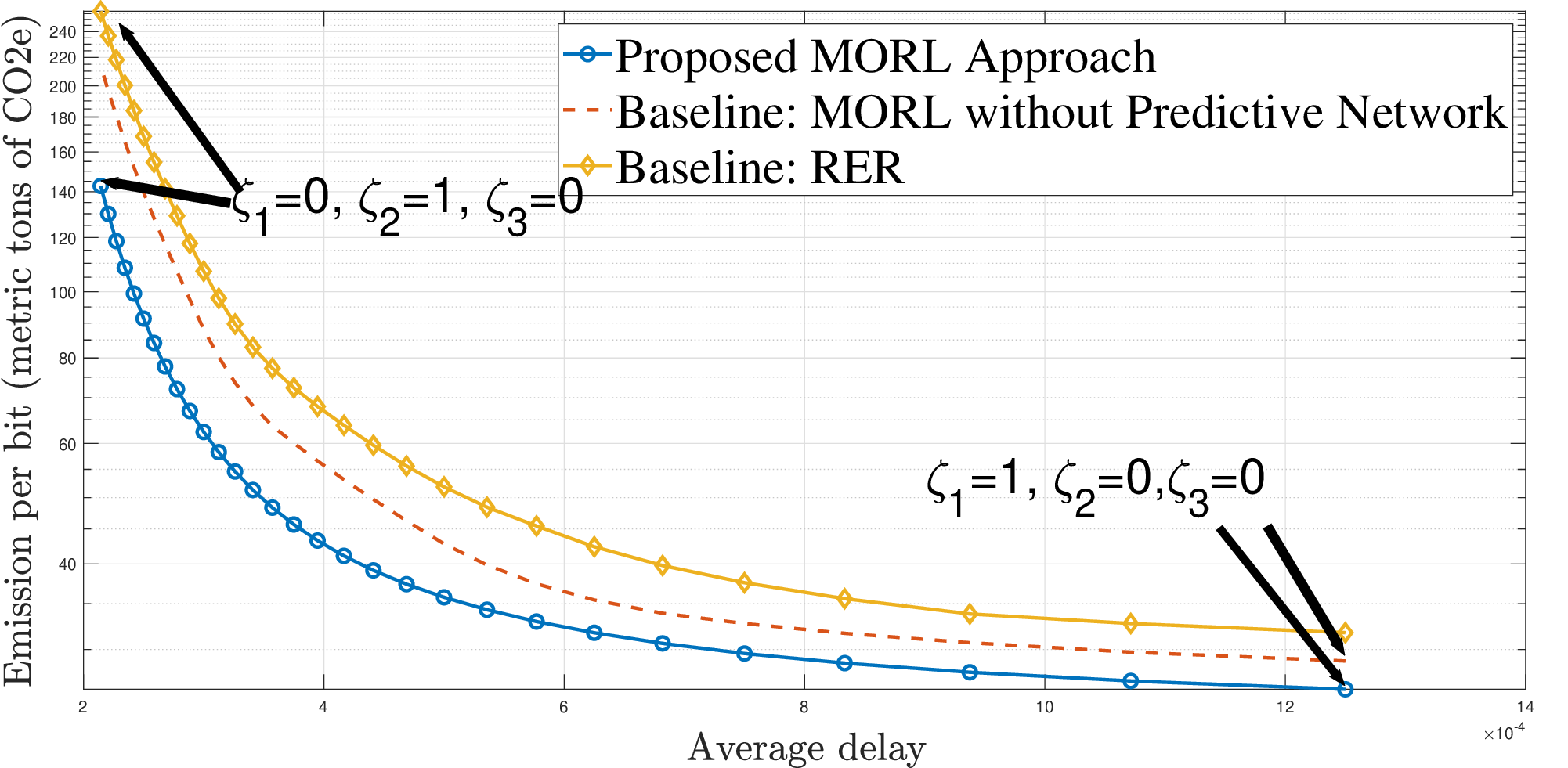}
\vspace{-2mm}\caption{\small Emissions per bit versus average delay tradeoff}
\label{fig:emissions_vs_delay} 
    \vspace{-8mm}
\end{figure}

Fig. \ref{fig:emissions_vs_delay} shows the tradeoff between the emissions per bit and the average delay. Depending on the weights $\zeta_r$ that control the relative importance of sustainability and average delay, the optimal global policy can vary. 
Our MORL approach achieves a $33\%$ reduction in emissions per bit at lower delay values (where the computational load is higher), with the reduction narrowing to $9\%$ at higher delay values. 
\vspace{-2mm}\section{Conclusion}\vspace{-1mm}

In this paper, we have introduced a novel sustainability measure  that quantifies the reduction in emissions per bit, considering the combined energy consumption for computing and communication, sourced from both renewable and non-renewable energy. We have formulated a MO optimization problem that simultaneously maximizes the average quality of service. We have devised a novel methodology using MORL to solve the resulting problem. Simulation results show that optimizing the proposed sustainability metric, which accounts for emissions per bit, enables efficient allocation of energy, computing, and communication resources, while achieving lower emissions per bit compared to state-of-the-art systems.

\bibliographystyle{IEEEbib}
\def\baselinestretch{0.9}
\vspace{-1mm}
\bibliography{refs,semantics}
\end{document}

%% file: defines.tex
\newcommand{\beq}{\begin{equation}}
\newcommand{\eeq}{\end{equation}}

\DeclareMathOperator*{\argmax}{arg\,max}

\def\adots{\mathinner{\mskip0mu\raise0pt\vbox{\kern7pt\hbox{.}}\mskip3mu
          \raise4pt\hbox{.}\mskip3mu\raise8pt\hbox{.}\mskip0mu}}

\usepackage{bm}

\newcommand{\bmx}{{\bm x}}

\newcommand{\mbE}{\mathbb{E}}

\newcommand{\mP}{\mathcal{P}}
\newcommand{\mD}{\mathcal{D}}

\newcommand{\mV}{\mathcal{V}}
\newcommand{\mN}{\mathcal{N}}

\newcommand{\mM}{\mathcal{M}}

\newcommand{\bmC}{{\bf C}}

\newcommand{\bma}{{\bm a}}